\documentclass[pra,twocolumn,showpacs,amssymb,aps,floatfix]{revtex4}
\newcommand{\be}{\begin{equation}}
\newcommand{\ee}{\end{equation}}
\newcommand{\bea}{\begin{eqnarray}}
\newcommand{\eea}{\end{eqnarray}}
\usepackage{graphicx}
\usepackage{amsmath}
\usepackage{color}

\begin{document}

\title{Generation and spectroscopic signatures of a fractional quantum Hall liquid of photons in an
incoherently pumped optical cavity}
\author{R. O. Umucal\i lar}
\affiliation{Department of Physics, Mimar Sinan Fine Arts University, 34380 Sisli, 
Istanbul, Turkey}
\author{I. Carusotto}
\affiliation{INO-CNR BEC Center and Dipartimento di Fisica, Universit\`a di Trento, I-38123 Povo, Italy}

\date{\today}
\begin{abstract}
We theoretically investigate a driven-dissipative model of strongly interacting photons in a nonlinear optical cavity in the presence of a synthetic magnetic field. We show the possibility of using a frequency-dependent incoherent pump to create a strongly-correlated $\nu = 1/2$ bosonic Laughlin state of light: thanks to the incompressibility of the Laughlin state, fluctuations in the total particle number and excitation of edge modes can be tamed by imposing a suitable external potential profile for photons. We further propose angular momentum-selective spectroscopy of the emitted light as a tool to obtain unambiguous signatures of the microscopic physics of the quantum Hall liquid of light.
\end{abstract}

\pacs{42.50.Pq, 42.50.Ar, 42.50.Ct, 73.43.−f}

\maketitle

\section{INTRODUCTION}

Simulation of a quantum system with another one which is thought to be easier to manipulate has become a popular research theme in the last two decades or so. This theme has started with advances in the ultracold-atom physics \cite{cold atoms} and  in the last few years has permeated the field of photonics \cite{light reviews}. Optical lattices for ultracold atoms and coupled optical cavities in photonics have appeared as analogues of solid-state lattices and synthetic magnetism for neutral particles has been intensely explored for both atoms \cite{synthetic field atom} and photons \cite{synthetic field photon}. Analogues of integer quantum Hall edge states have already been observed in photonic systems \cite{photonic edge states}. Nowadays one of the main goals of researchers studying synthetic magnetism is to create and manipulate strongly-correlated fractional quantum Hall (FQH) states of neutral particles both for the purpose of a better understanding of the underlying microscopic physics and for practical purposes like possible implementations in topological quantum computation~\cite{das sarma}. 

In the present work, we are interested in the possibility of creating and observing the $\nu = 1/2$ bosonic Laughlin state of FQH physics with a well-defined number of photons. We focus our attention on the non-planar ring cavity set-up of \cite{synthetic Landau levels}, where a strong synthetic magnetic field naturally appears from parallel transport along the non-planar optical path and the strength of the transverse harmonic confinement can be tailored via the mirror curvature. In contrast to the lattice geometries of~\cite{photonic edge states}, the multimode dynamics takes place in this system along the continuous two-dimensional transverse plane and has already revealed photonic Landau levels in the synthetic magnetic field. Strong photon-photon effective interactions can then be induced by coupling photons to atomic transitions in the so-called Rydberg EIT configuration \cite{Rydberg polariton expts}: the strong interactions between Rydberg atoms translate in analogous interactions between polaritons \cite{Rydberg polaritons} and first experimental evidence of photon blockade in the lowest mode has been recently reported in \cite{blockade}.

As the photon inside an optical cavity has a finite life time, in order to maintain a stable population of a correlated state of photons, one also needs to devise a method to compensate for photon losses. Previous works have focused on coherent drive mechanisms \cite{coherent Laughlin 1,coherent Laughlin 2,coherent Laughlin 3}, which look like a very simple and promising strategy to create Laughlin states of few photons. However, the efficiency of this approach is expected to quickly decrease for larger photon numbers as they employ multi-photon transitions with low probabilities and generally lead to a superposition of states with different particle numbers. Recently, new strategies have been put forward to remedy these unwanted properties of coherent pumping, including a flux insertion technique supplemented again by a coherent drive \cite{topological growing}. 

In the present work, we rely on a different pumping scheme based on a frequency-selective incoherent drive as proposed in \cite{induced stabilization,incoherent pumping} and further investigated in \cite{Biella,Lebreuilly2} in view of preparing correlated many-body states with a well-defined particle number, e.g. Mott insulators. While previous works in the quantum Hall framework~\cite{induced stabilization} concentrate on spatially homogeneous geometries with no boundaries, we here introduce an external potential profile \cite{external potential,elia} as a way to control the number of particles in the Laughlin state and suppress its excitation in state-of-the-art configurations~\cite{synthetic Landau levels}. For a given lateral confinement, the incompressibility of the Laughlin state puts a bound on the maximum number of particles that the frequency-dependent incoherent pump can inject into the cavity before the Laughlin gap pushes the transition off-resonance. The same confinement potential is also responsible for the splitting of the massive degeneracy of excitations of the Laughlin state \cite{elia} and therefore serves to suppress quasi-hole and edge excitations.

As a second remarkable result, we show how information on the microscopic physics of the quantum Hall liquid of light can be extracted from an angular-momentum-resolved spectroscopy of the light emitted by the cavity: Taking inspiration from a related proposal for rotating atomic gases \cite{Cooper multiplicity}, we show how clear signatures of the fractional exclusion statistics can be obtained by looking at the spectral distribution of the emission lines and at their unique multiplicity structure.

The paper is organized as follows. Section~\ref{sec:model} describes our model in two parts, focusing on the lossless isolated system in the first part and detailing the incoherent pumping mechanism in the second. Our numerical results are presented in Section~\ref{sec:preparation}, starting with the possibility of populating a target Laughlin state with a well-defined particle number by changing the parameters of the external potential. Section~\ref{sec:detection} then presents a discussion of the multiplicity structure of the allowed transitions and illustrates our predictions about the possibility of extracting information on the Laughlin physics from measurements of the spectral and coherence properties of the emitted light. The final Section~\ref{sec:conclu} is devoted to conclusions. Technical details on the Laughlin state preparation, on the external potential, and on optical transitions from a Laughlin state are given in the Appendices.

\section{The physical system and the theoretical model}
\label{sec:model}

In the first part of this section, we review the theoretical description of the isolated cavity dynamics. In the second part, we then deal with pumping and losses and introduce the incoherent pumping mechanism used to maintain a stable population of the target FQH state. 

\subsection{Isolated system}

We consider an optical set-up inspired by the recent experiment~\cite{synthetic Landau levels}, namely a non-planar ring cavity. Assuming that the free spectral range of the cavity is larger than all other energy scales of the photon fluid, we can assume that photons only occupy a single running-wave longitudinal mode of the ring cavity and their motion is confined to the two-dimensional $xy$ transverse plane. The mirrors are assumed to have cylindrical symmetry around their axes, so that they provide a tunable effective harmonic confinement. An additional potential can be imposed to the photons by means of an $xy$-dependent phase shifter element. The non-planar shape of the cavity gives rise to a uniform synthetic magnetic field along the perpendicular direction $z$. Finally, for the sake of simplicity we suppose that interactions between photons can be modelled as binary contact interactions. While this is generally accurate for semiconductor materials~\cite{light reviews}, corrections due to the long-range nature of the atom-atom interactions might be needed in the Rydberg-EIT case \cite{Rydberg polaritons}, especially at relatively high photon densities. A detailed analysis of these effects goes beyond the present article and will be the subject of future work.

In formal terms, this system can be described in terms of the following second-quantized Hamiltonian involving a two-dimensional bosonic field operator $\Psi(\mathbf{r})$: 
\begin{multline}
\mathcal{H}
=\int\!d^2\mathbf{r}\,\left\{\Psi^\dagger(\mathbf{r})\left[\frac{(-i\hbar\nabla-\mathbf{A}(\mathbf{r}))^2}{2m_{ph}}\,+\hbar\omega_{cav}\,+ \right.\right.\\ \left.\left.+V_{ext}(r)\right]\Psi(\mathbf{r})+\frac{\hbar g_{nl}}{2}\,\Psi^\dagger(\mathbf{r})\Psi^\dagger(\mathbf{r})
\Psi(\mathbf{r})
\Psi(\mathbf{r})\right\}.
\label{eq:H}
\end{multline}
In this Hamiltonian, terms in square brackets correspond to the single particle Hamiltonian: in particular, $\omega_{cav}$ is the natural frequency of the longitudinal mode under consideration and $m_{ph} = \hbar\omega_{cav}/c^2$ is the effective photon mass that results from confinement along $z$. The cylindrically symmetric $V_{ext}(r)$ describes the total external confinement potential.

The synthetic magnetic vector potential $\mathbf{A}(\mathbf{r})$ is taken to have the symmetric-gauge form $\mathbf{A} = B\hat{\mathbf{z}}\times\mathbf{r}/2$ corresponding to a uniform perpendicular synthetic magnetic field $\mathbf{B} = \mathbf{\nabla}\times\mathbf{A} = B\hat{\mathbf{z}}$ acting on photons of unit synthetic charge. The last term in Eq. (\ref{eq:H}) is the interaction Hamiltonian $\mathcal{H}_{int}$ describing repulsive contact interactions between photons with strength $g_{nl}$, which is determined by the properties of the underlying nonlinear medium.

\begin{figure}[htbp]
\includegraphics[scale=0.3]{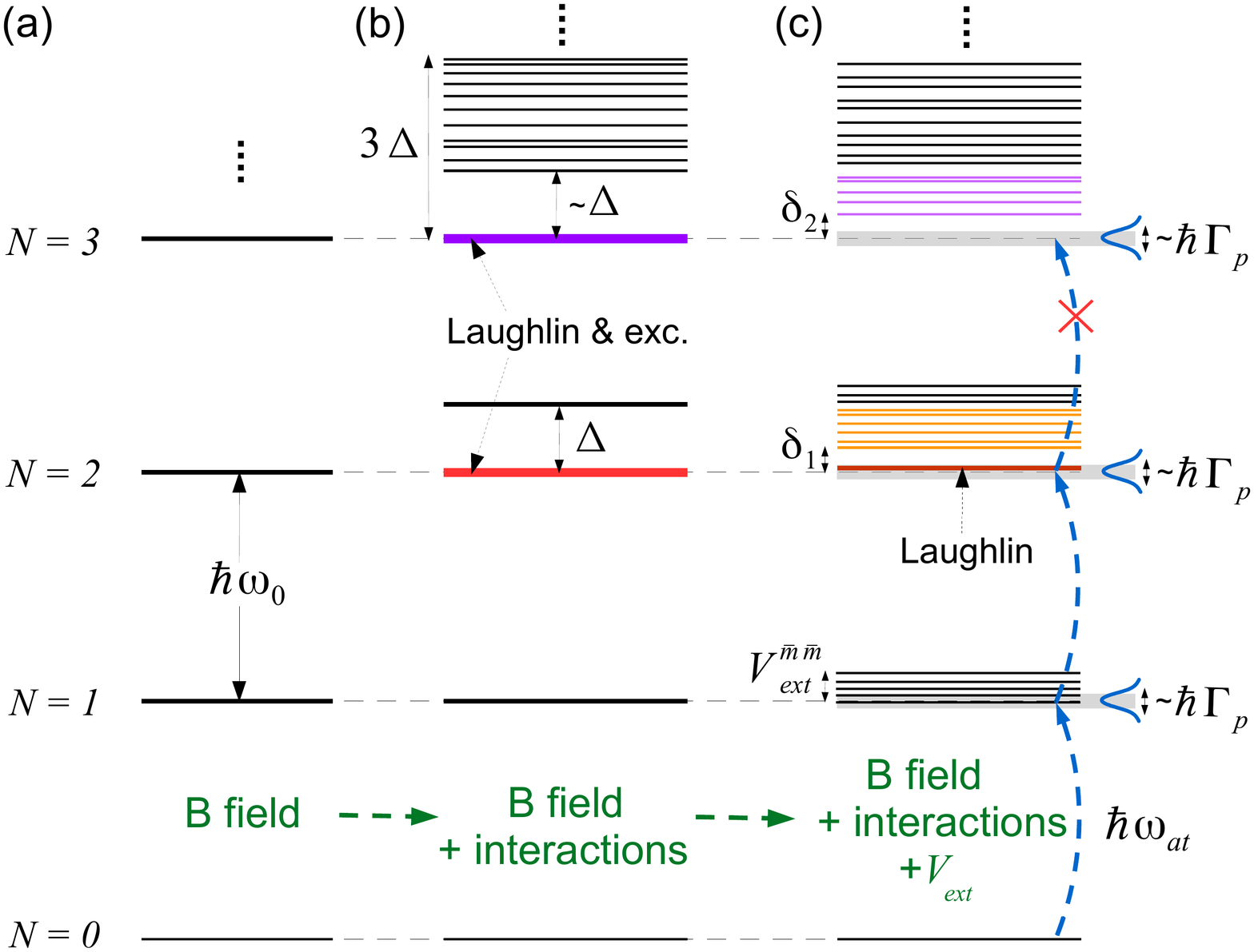}
\caption{Schematic energy levels of the isolated system. (a) In the presence of a magnetic field, noninteracting particles occupy the lowest Landau level. The energy separation between each successive $N$-particle manifold is $\hbar\omega_0$. (b) When interactions are included, the ground state for each $N$ is massively degenerate and contains the $\nu = 1/2$ bosonic Laughlin state and its zero-energy edge and quasi-hole excitations (shown in red for $N = 2$, purple for $N = 3$). This Laughlin manifold is separated from all other excitations by a gap $\sim \Delta$. (c) Low-energy states in the presence of an external potential with matrix elements $V_{ext}^{\bar{m}\bar{m}}$. Degeneracies are lifted and for a suitably designed potential profile it is possible to appreciably blueshift all the excitations (shown in orange) of a target $N$-particle Laughlin state and all energies in the $(N+1)$-particle Laughlin manifold (shown in light purple). In both these cases, states above the Laughlin gap $\Delta$ are marked in black. If the transition frequency $\omega_{at}$ of two-level emitters matches $\omega_0$, the target Laughlin state (here $N=2$) can be sizably populated provided the energy shifts $\delta_1,\delta_2,\ldots$ are larger than the pump linewidth $\Gamma_p$.   
\label{fqh_levels_schematic}}
\end{figure}

In contrast to other experimental studies of quantum Hall effect with light \cite{photonic edge states}, our chosen configuration corresponds to a two-dimensional continuum model whose single particle eigenstates in the absence of external potential reduce to Landau levels with equally-spaced energies separated by $2\hbar\omega_{cycl} = \hbar B/m_{ph}$. The energy of a photon in the lowest Landau level (LLL) is then $\hbar\omega_0\equiv \hbar(\omega_{cycl}+\omega_{cav})$, so the total energy of $N$ noninteracting photons in the LLL is simply given by $E^{(0)}_N = N\hbar\omega_0$ [Fig.~\ref{fqh_levels_schematic}(a)]. In the following, we shall use this $\omega_0$ as a reference point for the optical frequencies.

In the rotationally symmetric gauge that we are using, single particle states can be classified in terms of their angular momentum. The wave function of a single-particle eigenstate in the LLL with angular momentum $m\hbar$ has the simple form $\varphi_m(z) = z^{m}e^{-|z|^{2}/2}/\sqrt{\pi m!}$, where $z = (x + iy)/\ell$ is the complex-valued coordinate of the particle and $\ell = \sqrt{\hbar/m_{ph}\omega_{cycl}}$ is the magnetic length. When the typical interaction energy represented by the lowest Haldane pseudo-potential $v_0 = \hbar g_{nl}/2\pi\ell^2$ for the contact interaction is much smaller than the separation between Landau levels, that is $v_0\ll \hbar\omega_0$, the low-energy physics is restricted to the LLL. 

In the absence of an external potential (${V}_{ext} = 0$), it is well-known that the ground-state of the interacting $N$-particle system is heavily degenerate [Fig.~\ref{fqh_levels_schematic}(b)]. The lowest-total-angular-momentum state in this degenerate manifold is the bosonic $\nu = 1/2$ Laughlin state of FQH physics \cite{Laughlin state}
\begin{equation}
\Psi_{\rm FQH}(z_1, \ldots, z_N) \propto \prod_{j<k}(z_j-z_k)^2e^{-\sum_{i = 1}^N|z_i|^2/2},\label{WF_FQH}
\end{equation}
where $z_j$ is the coordinate of the $j$th particle. This $\nu = 1/2$ Laughlin state has a total angular momentum $L_z = N(N-1)\hbar$ and is not degenerate with any other state with same total angular momentum. All other states in the degenerate manifold have a larger total angular momentum and correspond to zero-energy edge or quasi-hole excitations of the Laughlin state. Their wave function representation can be expressed as a product of the Laughlin wave function Eq. (\ref{WF_FQH}) and polynomials symmetric in the particle coordinates.
This lowest-energy Laughlin manifold is separated from the excited states by a gap $\sim \Delta$, where $\Delta=v_0$ is also the exact gap for two particles in the LLL approximation. In this limit, the highest-energy excitations are again degenerate states with pure center-of-mass motion, whose energy is greater than that of the Laughlin manifold by $N(N-1)\Delta/2$ [Fig.~\ref{fqh_levels_schematic}(b)] \cite{coherent Laughlin 3}.

\begin{figure}[htbp]
\includegraphics[scale=0.5]{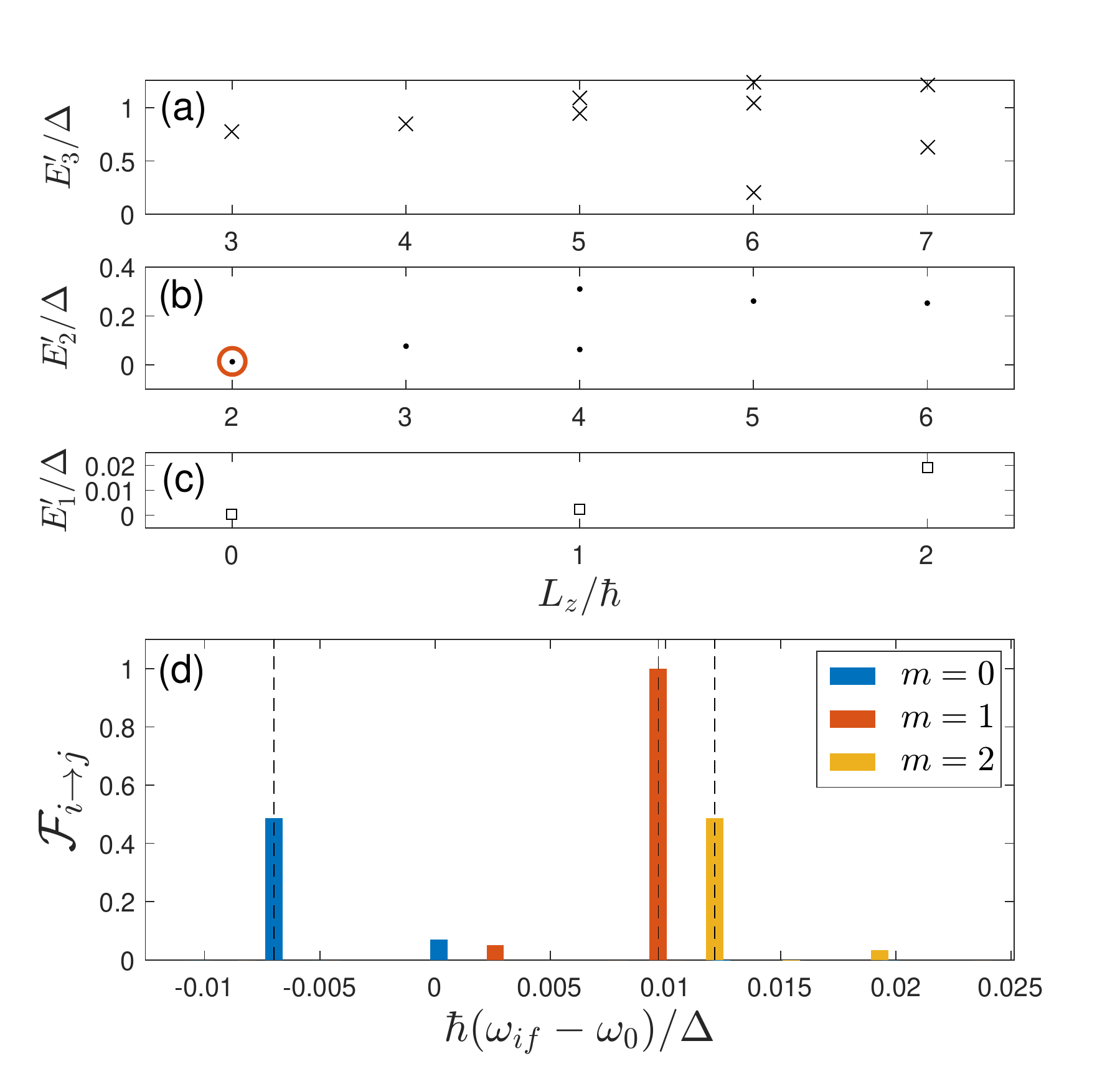}
\caption{Low-lying energy levels $E^{\prime}_N = E_N-N\hbar\omega_0$ of the isolated system in the presence of the step potential with $V/\Delta = 50\pi$ and $R/\ell = 3.7$, versus total angular momentum $L_z$: (a) $N = 3$, (b) $N = 2$, (c) $N = 1$. (d) Scaled Fourier amplitudes $\mathcal{F}_{i\rightarrow f}^{(m)}$ of transitions as a function of the difference between the lost-photon frequency $\omega_{if} = \omega_i-\omega_f$ and the reference frequency $\omega_0$. Different colors designate the angular momentum $m\hbar$ of the photon lost from the cavity. Vertical dashed lines show the position of allowed transitions from the $N = 2$ Laughlin state [inside the red circle in (b)] to the three states in the $N = 1$ manifold of (c). Same loss and pump parameters as in Fig.~\ref{PopsVR}.
\label{FourierN2to1}}
\end{figure}

\begin{figure}[htbp]
\includegraphics[scale=0.5]{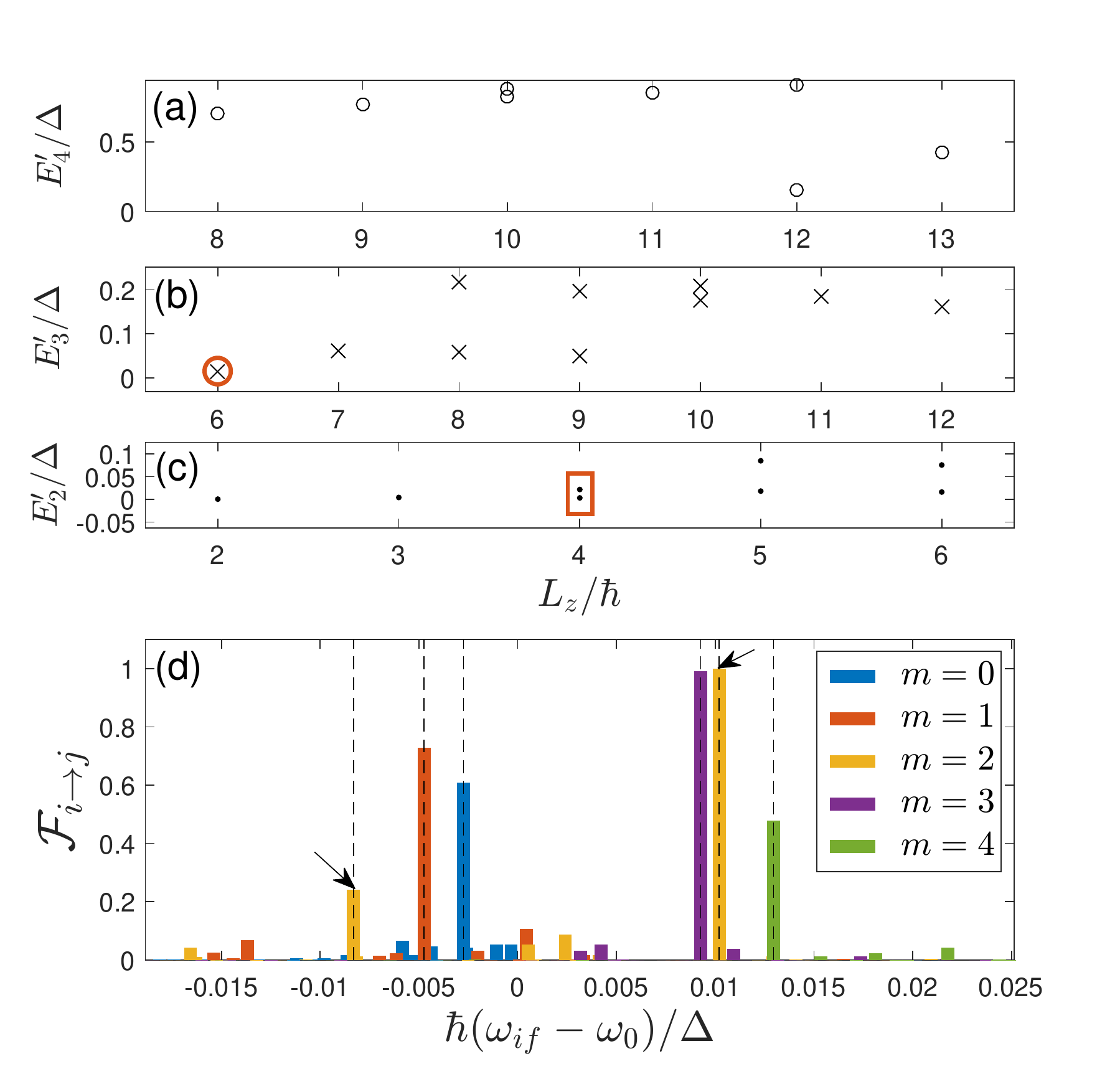}
\caption{Low-lying energy levels $E^{\prime}_N = E_N-N\hbar\omega_0$ of the isolated system in the presence of the step potential with $V/\Delta = 50\pi$ and $R/\ell = 4.15$, versus total angular momentum $L_z$: (a) $N = 4$, (b) $N = 3$, (c) $N = 2$. (d) Scaled Fourier amplitudes $\mathcal{F}_{i\rightarrow f}^{(m)}$ of transitions as a function of the difference between the lost-photon frequency $\omega_{if} = \omega_i-\omega_f$ and the reference frequency $\omega_0$. Different colors designate the angular momentum $m\hbar$ of the photon lost from the cavity. Vertical dashed lines show the position of allowed transitions from the $N = 3$ Laughlin state [inside the red circle in (b)] to the six lowest-energy states in the $N = 2$ manifold of (c). Also shown inside a red rectangle are the two states degenerate with the $N = 2$ Laughlin state in the absence of an external potential. Transitions to these two states with $m = 2$ are indicated by arrows in (d). Same loss and pump parameters as in Fig.~\ref{PopsVR}.
\label{FourierN3to2}}
\end{figure}

In Fig.~\ref{fqh_levels_schematic}(c), we schematically show the few lowest energy levels in different particle-number manifolds in the presence of an external potential. The first effect of such a potential is to lift the degeneracies in the Laughlin manifold: If the external potential is judiciously chosen, it is possible to shift all the edge and quasi-hole excitations of an N-particle Laughlin state by an amount $\delta_{N-1}$ [$\delta_1$ in Fig.~\ref{fqh_levels_schematic}(c)] on the order of $V_{ext}^{\bar{m}\bar{m}} \equiv \int d^2\mathbf{r}\varphi^{\ast}_{\bar{m}}(\mathbf{r})V_{ext}(r)\varphi_{\bar{m}}(\mathbf{r})$, while not shifting the energy of the $N$-particle Laughlin state appreciably. Here, the angular momentum $\bar{m}\hbar = [2(N-1)+1]\hbar$ value is defined to be one unit of angular momentum larger than the highest angular momentum orbital present in the $N$-particle Laughlin state. The $(N+1)$-particle Laughlin manifold is altogether shifted by an amount $\delta_N$ [$\delta_2$ in Fig.~\ref{fqh_levels_schematic}(c)] again on the order of $V_{ext}^{\bar{m}\bar{m}}$. 

A quantitative view on this physics can be found in Figs.~\ref{FourierN2to1}(a,b,c) and \ref{FourierN3to2} (a,b,c) for two exemplary potential choices. In these panels, we show the low-lying many-body energy levels as a function of angular momentum for different particle numbers. In both cases, a cylindrically symmetric step potential with a sudden jump of finite strength $V$ at a given radius $R$, $V_{ext}(r) = V\Theta(r-R)$ (where $\Theta$ is the Heaviside step function) is considered, with suitably chosen $V,R$ parameters in view of looking for the $N=2,3$ Laughlin states, respectively. In all panels, one can easily recognize the Laughlin manifold starting from angular momentum $N(N-1)\hbar$ and well separated from further excited states by a gap of order $\Delta$. 

Within the Laughlin manifold, the confinement potential has little effect on low-angular momentum states upto the desired Laughlin state, while it shifts upwards the excited states of this Laughlin manifold as well as all higher-$N$ Laughlin states, including the fundamental Laughlin state. For the considered potential choice, note however how the size of this confinement-induced energy shifts are smaller than $\Delta$. While too a strong confinement potential may overcome the Laughlin gap of order $\Delta$ and mix states in the Laughlin manifold with higher-lying excited ones, the improved forms of the potential discussed in Appendix B and in \cite{elia} may help having both energy scales of the same order.

As a suitably chosen confinement potential plays an essential role to remove the massive degeneracy of the Laughlin state and its excitations, and blueshift higher $N$ states, an active effort is presently devoted to its optimization in view of effectively populating the chosen $N$-particle Laughlin state and then detecting the signatures of FQH physics from the emission.

\subsection{Losses and Incoherent Pumping}
\label{subsec:Losses and Incoherent Pumping}

The basic idea underlying our proposal is to use a frequency-dependent incoherent pump so as to be able to excite the system up to the desired $N$-particle Lauglin state, but then block any further excitation to higher $N$'s or to excited $N$-particle states. A pictorial representation of our idea is illustrated in Fig.\ref{fqh_levels_schematic}(c).

As it is discussed in detail in~\cite{incoherent pumping} and the Appendices therein, this can be accomplished by placing in the cavity a large number of population-inverted two-level emitters of transition frequency $\omega_{at}$, in order to obtain an effectively Lorenzian emission spectrum. More precisely, we assume that the emitters are quickly pumped to their excited state at a rate $\Gamma_p$ much larger than the collective Rabi frequency $\Omega_R$ of the collective cavity field-emitter coupling. Under this condition, the emitters do not have time to reabsorb a cavity photon before being pumped back to their excited state. Provided the spontaneous decay rate $\gamma$ is also much smaller than $\Gamma_p$, the emitters are found for most of the time in their excited state and it is possible to trace out the emitter degrees of freedom and write a closed master equation for the density matrix $\rho$ of cavity photons only.

This master equation turns out to be composed of three parts
\bea
\frac{\partial \rho}{\partial t}\ = -\frac{i}{\hbar}[\mathcal{H},\rho]+\mathcal{L}_l+\mathcal{L}_e.
\label{Master}
\eea
The first commutator term corresponds to the usual unitary evolution of the photonic Hamiltonian $\mathcal{H}$.
The frequency-independent photonic losses of rate $\Gamma_l$ are described by the standard Lindblad superoperator
\begin{multline}
\mathcal{L}_l= \frac{\Gamma_l}{2}\int\!d^2\mathbf{r}\,\left[2\Psi(\mathbf{r}) \rho \Psi^\dagger(\mathbf{r})\right.\\
\left.-\Psi^\dagger(\mathbf{r})\Psi(\mathbf{r})\rho - \rho\Psi^\dagger(\mathbf{r})\Psi(\mathbf{r})\right].
\label{Loss}
\end{multline}
The frequency selectivity of the emission processes requires instead a more sophisticated description in terms of a generalized superoperator of the form
\begin{multline}
\mathcal{L}_e= \frac{g_e}{2}\int\!d^2\mathbf{r}\,n_{\rm at}(\mathbf{r})\,\left[\tilde{\Psi}^\dagger(\mathbf{r}) \rho \Psi(\mathbf{r})+\Psi^\dagger(\mathbf{r}) \rho \tilde{\Psi}(\mathbf{r})\right.\\
\left.-\Psi(\mathbf{r})\tilde{\Psi}^\dagger(\mathbf{r})\rho - \rho\tilde{\Psi}(\mathbf{r})\Psi^\dagger(\mathbf{r})\right].
\label{Emission}
\end{multline}
Here, $n_{\rm at}(\mathbf{r})$ is the atomic density at position $\mathbf{r}$ and $g_e = 4\,\omega_{cav}\,|d_{eg}|^2/(\hbar\Gamma_p L_{\perp})$ quantifies the coupling of each atom to radiation in terms of the dipole matrix element $d_{eg}$, the cavity length $L_{\perp}$ and the repumping rate $\Gamma_p$.  In the following, we shall assume that the atomic density is constant $n_{\rm at}(\mathbf{r})=n_{\rm at}$ in the spatial region of interest.

In the formalism of \cite{incoherent pumping}, this frequency selectivity is included by the modified field operator $\tilde{\Psi}(\mathbf{r})$ defined as
\begin{equation}
\tilde{\Psi}(\mathbf{r}) = \frac{\Gamma_p}{2}\int_0^{\infty}\!d\tau\, e^{(-i\omega_{at}-\Gamma_p/2)\tau}\Psi(\mathbf{r},-\tau),
\label{Modified field}
\end{equation}
where we have assumed that loss and emission rates $\Gamma_l$ and $\Gamma_e$ are much slower than the repumping rate $\Gamma_p$. The latter condition $\Gamma_e\ll \Gamma_p$ automatically follows from the initial hypothesis $\Omega_R\ll \Gamma_p$ needed to exclude collective Rabi oscillation in favor of an irreversible emission process. Under these same approximations, we can finally write
\begin{equation}
\Psi(\mathbf{r},\tau) \simeq e^{i\mathcal{H}\tau/\hbar}\Psi(\mathbf{r})e^{-i\mathcal{H}\tau/\hbar}
\end{equation}
which only depends on the Hamiltonian evolution.

Using the Hermitian conjugate of Eq.~(\ref{Modified field}) directly, the matrix elements of the modified field operator in the eigenstate basis of $\mathcal{H}$ can be written in terms of the original ones as
\begin{multline}
\langle N+1|\tilde{\Psi}^\dagger(\mathbf{r})|N\rangle = \frac{\Gamma_p/2}{-i(\omega_{at}-\omega_{N+1,N})+\Gamma_p/2}\\
\times\langle N+1|\Psi^\dagger(\mathbf{r})|N\rangle,
\label{Modified original relation}
\end{multline}
where $|N\rangle$ ($|N+1\rangle$) is an $N$ ($N+1$)-particle eigenstate of $\mathcal{H}$ and $\omega_{N+1,N} \equiv (E_{N+1}-E_N)/\hbar$ is the difference between the eigenfrequencies. The real part of the factor connecting the original and modified matrix elements in Eq.~(\ref{Modified original relation}) sets the effective emission rate for each $|N\rangle \rightarrow |N+1\rangle$ transition. This rate attains its maximum value $\Gamma_e=g_e n_{\rm at}$ when the emitter transition frequency $\omega_{at}$ matches the separation $\omega_{N+1,N}$ between the frequencies of two many-particle states. Off-resonance, the emission rate is suppressed according to a Lorentzian lineshape of linewidth $\Gamma_p$, which provides the desired frequency-selectivity of the emission.

Returning back to Laughlin physics in the presence of an external potential, we see that the incoherent pumping mechanism is very efficient in populating a particular $N$-particle Laughlin state provided that the emitter frequency $\omega_{at}$ is chosen to match the frequency interval between pairs of unperturbed (or slightly perturbed) states with successive number of particles, roughly constant and equal to $\omega_0$ from vacuum up to the desired $N$-particle Laughlin state. Assuming that losses are much slower than the resonant emission, $\Gamma_e\gg\Gamma_l$, this choice of $\omega_{at}$ makes in fact the emission process to dominate over losses as long as emission can be resonant. 

The cutoff in the particle number needed to focus on a particular $N$-particle Laughlin state is introduced by the external potential which blueshifts the energy of all states with $N^{\prime}>N$ particles. If this blueshift is sufficiently larger than the pump linewidth $\Gamma_p$, the emission into the blueshifted states is effectively blocked and we expect a very large population to accumulate in the $N$-particle state. A similar mechanism is effective in blocking all higher-total-angular-momentum states with $N$ particles that were initially degenerate with the target Laughlin state. A sketch of this mechanism is given in Fig.~\ref{fqh_levels_schematic}(c): when $\delta_1,\delta_2\gg\Gamma_p$, the population will mostly be concentrated in the $N = 2$ Laughlin state. A similar mechanism is at work for target states with higher $N$.

As compared with our previous coherent pumping proposal~\cite{coherent Laughlin 1}, it is useful to note how pumping to the $N$-particle Laughlin state occurs here via a sequence of resonant intermediate states belonging to the $N'<N$ particle Laughlin manifold, while the spectral separation of the interacting excited states guarantees that their actual population is strongly suppressed. As a result, the peculiar quantum correlations of the final non-interacting Laughlin state are progressively built by adding one particle at a time.

\section{Laughlin state preparation}
\label{sec:preparation}

After having introduced the general framework, in this section we present our results obtained from a numerical solution of the master equation (\ref{Master}), so as to validate the efficiency of the proposed frequency-dependent incoherent pumping scheme to generate fractional quantum Hall states of light. 

We used a superoperator method \cite{superoperator} to find the steady-state density matrix $\rho_{ss}$. The block diagonal form of $\rho_{ss}$ in the particle number and in the total angular momentum basis $|N,L_z\rangle$ was exploited to reduce the dimension of the superoperator. Due to computational difficulties arising from the huge dimension of the superoperator, we set $N = 4$ as the maximum particle number in our calculations and focused our attention on the $N = 2,3$ Laughlin states. We are however confident that our results should straightforwardly extend to larger number of particles once a suitable external potential profile is applied. 

In all calculations presented in this section, we chose $\hbar\Gamma_l/\Delta = \pi\times10^{-5}$, $\hbar\Gamma_e/\Delta = 5\pi\times10^{-4}$ and $\Gamma_p = 10\Gamma_e$ in accordance with the requirement $\Gamma_e \ll \Gamma_p$ discussed in Section \ref{subsec:Losses and Incoherent Pumping} (cf. Appendix A for the choice of parameters). For simplicity we consider a cylindrically symmetric step potential with a sudden jump of finite strength $V$ at a given radius $R$, $V_{ext}(r) = V\Theta(r-R)$, where $\Theta$ is the Heaviside step function. Examples of many-body spectra for different particle numbers and different angular momenta are shown in Figs.~\ref{FourierN2to1}(a,b,c) and \ref{FourierN3to2}(a,b,c). The effect of choosing different potential profiles is briefly discussed in Appendix B.

\begin{figure}[htbp]
\includegraphics[scale=0.057]{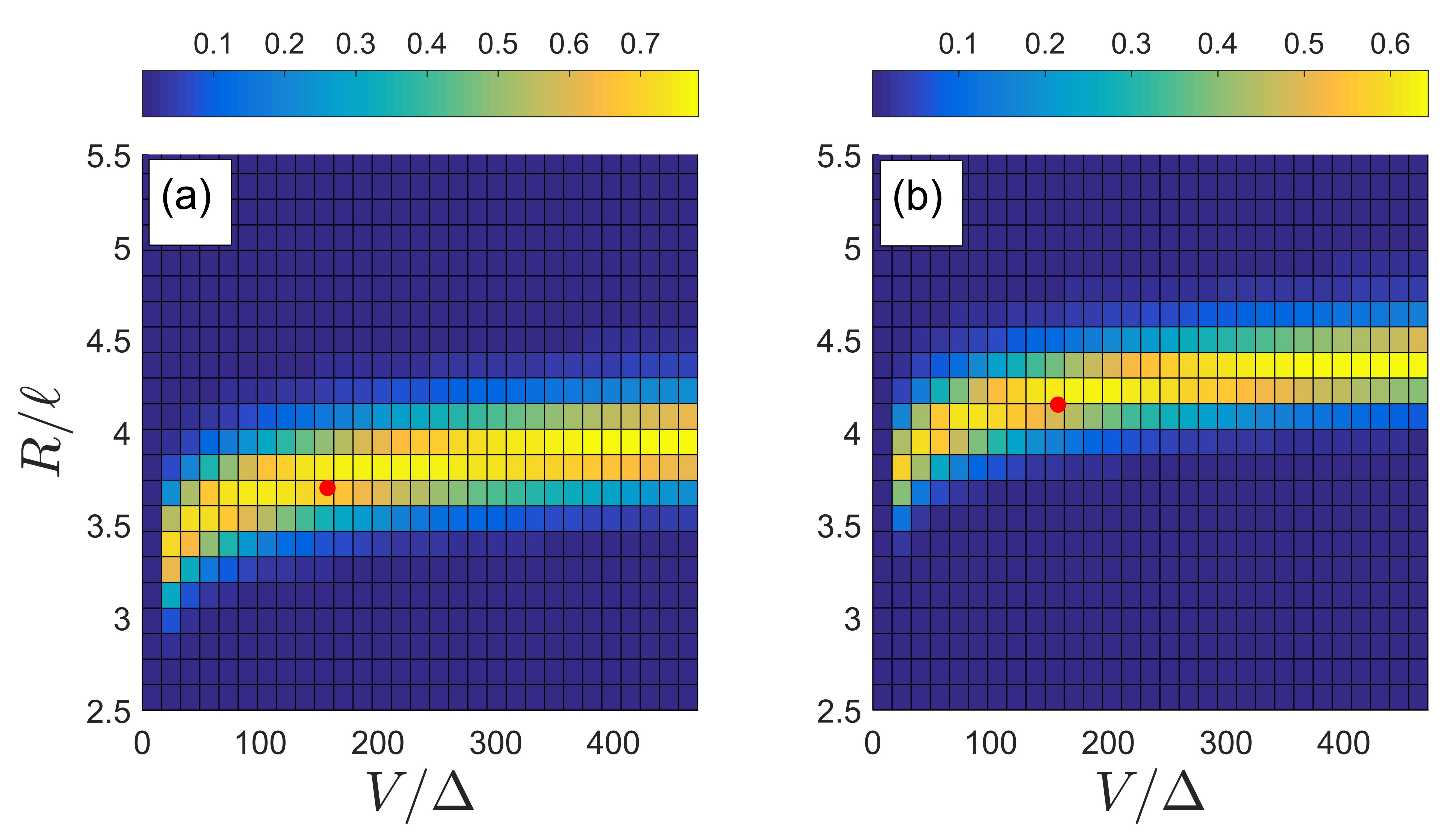}
\caption{Laughlin state population $P^{(N)}_{\rm FQH}$ as a function of the strength $V$ and the radius $R$ of the cylindrically-symmetric step-potential, $V_{ext}(r) = V\Theta(r-R)$ for different target particle numbers, $N = 2$ (a) and $N = 3$ (b). Red dots correspond to the $V$ and $R$ values used for subsequent results. Loss and incoherent-pumping parameters: $\hbar\Gamma_l/\Delta = \pi\times10^{-5}$, $\hbar\Gamma_e/\Delta = 5\pi\times10^{-4}$, $\hbar\Gamma_p/\Delta = 5\pi\times10^{-3}$, $\omega_{at}=\omega_0$. \label{PopsVR}}
\end{figure}

In Fig.~\ref{PopsVR} we show the $N$-particle Laughlin state population $P_{\rm FQH}^{(N)} \equiv \langle\Psi_{\rm FQH}^{(N)}|\rho_{ss}|\Psi_{\rm FQH}^{(N)}\rangle$ for  $N = 2,3$ as a function of $V$ and $R$ for the fixed value of $\omega_{at} = \omega_0$ and for the other pump parameter values cited above. These are chosen in a way to have $\Gamma_p$ significantly smaller than the Lauglin gap as well as of the confinement-induced blueshifts. Recall that the microscopic origin of the non-Markovianity of our pumping process imposes that the losses $\Gamma_l$ and the resulting emission rate $\Gamma_e$ are in turn smaller than $\Gamma_p$. 

From the plots, it is apparent that the population of the Laughlin state becomes appreciable only in a narrow band, but can reach quite important values of $0.78$ for $N = 2$ [panel (a)] and $0.64$ for $N = 3$ [panel (b)]. As usual, the unit trace condition on the density matrix imposes that $\sum_N P^{(N)}=1$, the $P^{(N)}$'s being the total population of $N$-particle states. From the graphs, one can also see how the narrow band with relatively large $P_{\rm FQH}^{(N)}$ tends to shift towards larger values of the radius $R$ as the particle number $N$ is increased from $N=2$ to $N=3$: this is quite expected from the incompressibility of the Laughlin state, which requires a larger space to accommodate more particles. 

Remarkably, one can extract from our simulations that the optimal condition for $P_{\rm FQH}^{(N)}$ corresponds to a potential profile for which the matrix elements $V_{ext}^{mm}$ do not vary appreciably upto and including the largest angular momentum $\bar{m}\hbar = 2(N-1)\hbar$ orbital occupied in the Laughlin state, but rapidly increase starting from $\bar{m}+1$. Potential profiles of such form not only have the effect of blueshifting all states in the $(N+1)$-particle Laughlin manifold, but also of isolating in energy the $N$-particle Laughlin state from its edge and quasi-hole excitations, which contain particles in higher angular momentum orbitals. An improved potential profile is discussed in Appendix B, but the improvement does not turn out to be dramatic. Further studies in this direction are in progress.

\begin{figure}[htbp]
\includegraphics[scale=0.07]{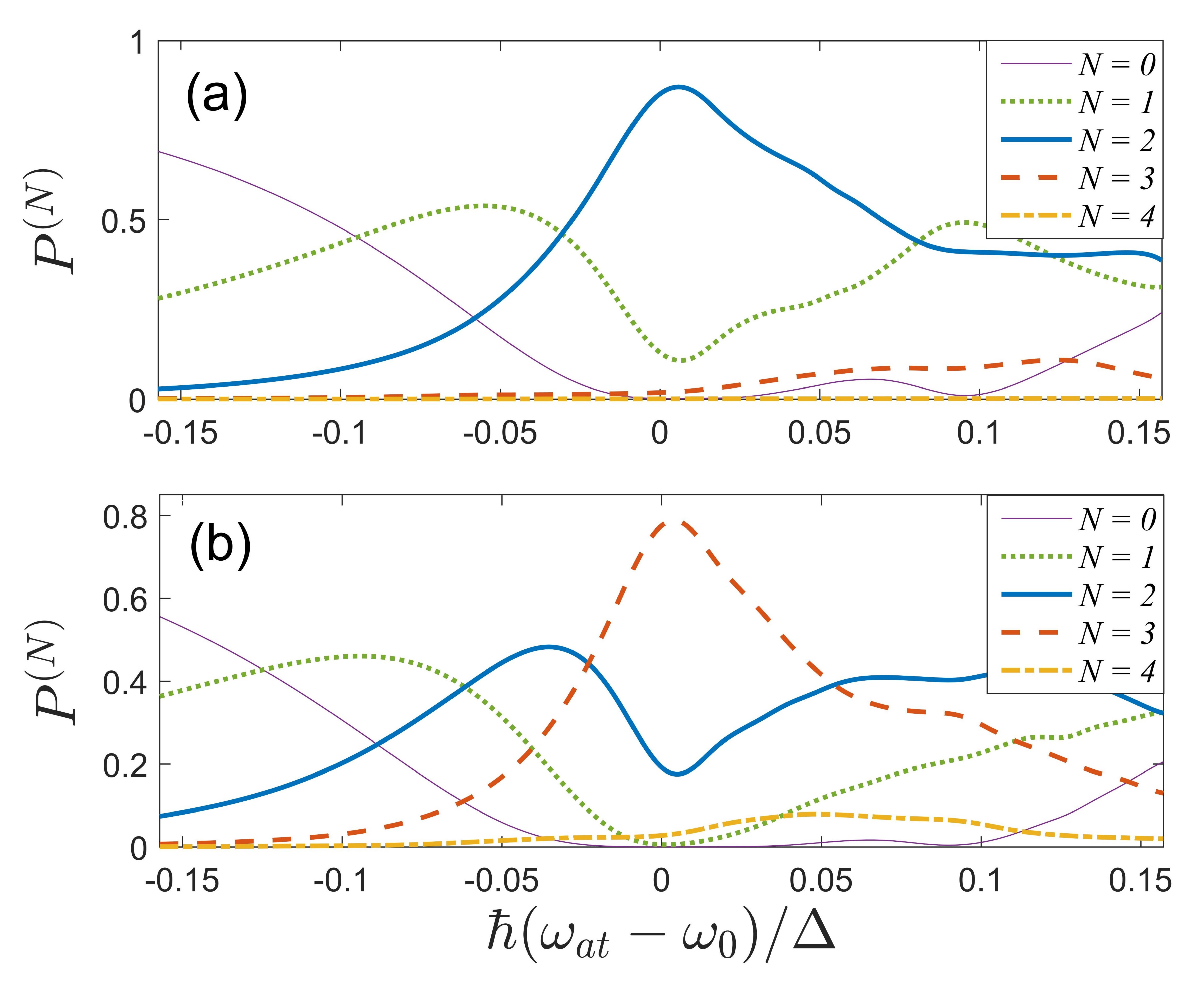}
\caption{Total population $P^{(N)}$ of the different $N$-particle states as a function of the detuning $\omega_{at}-\omega_0$ for step-potential parameters $V/\Delta = 50\pi$ and (a) $R/\ell = 3.7$, (b) $R/\ell = 4.15$ chosen to obtain a $N=2$ or $N=3$ target Laughlin state. Same loss and pump rates as in Fig.~\ref{PopsVR}.   
\label{Populations_w_at}}
\end{figure}

\begin{figure}[htbp]
\includegraphics[scale=0.07]{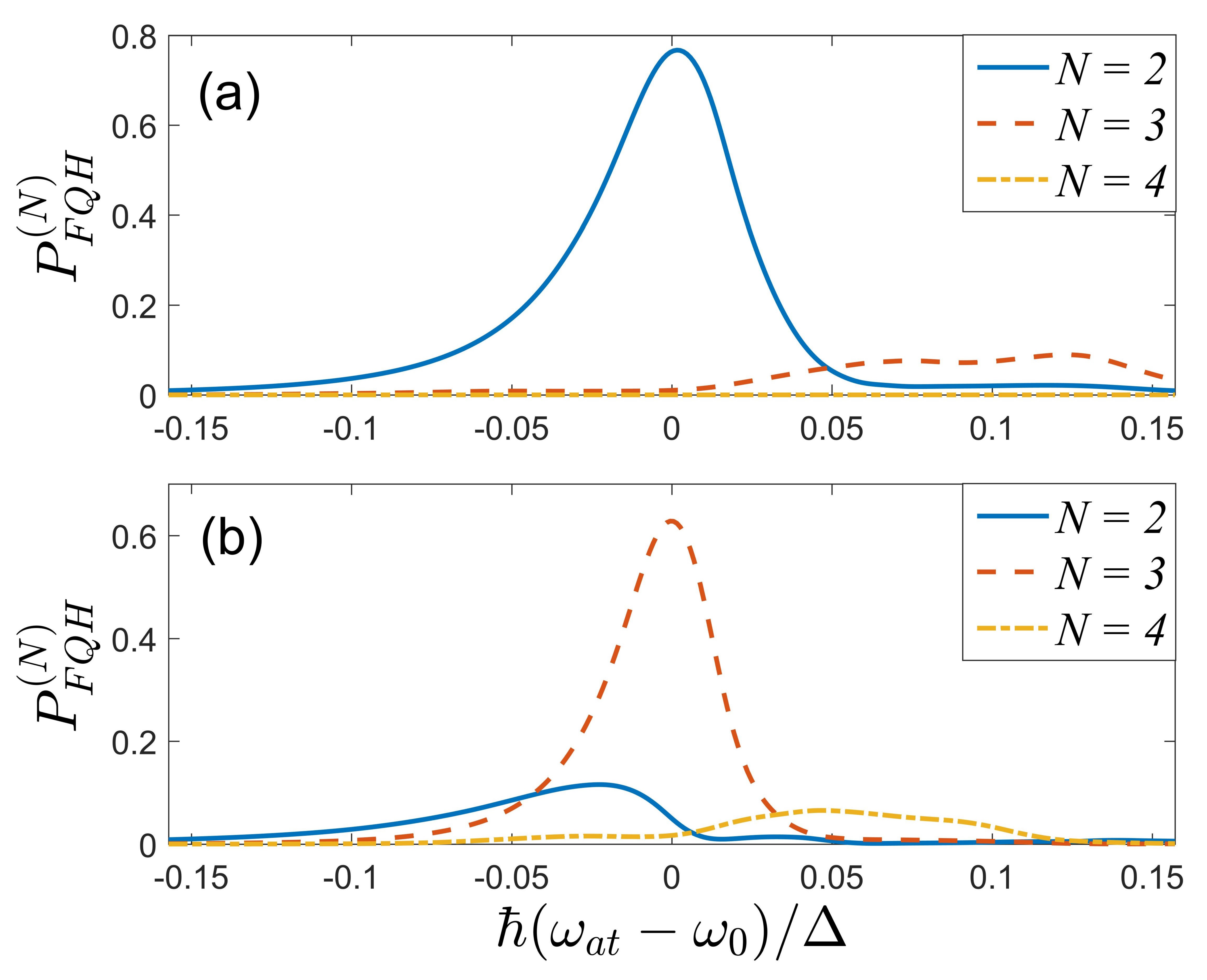}
\caption{$N$-particle Laughlin state populations $P^{(N)}_{\rm FQH}$ as a function of the detuning $\omega_{at}-\omega_0$  for step-potential parameters $V/\Delta = 50\pi$ and (a) $R/\ell = 3.7$, (b) $R/\ell = 4.15$ chosen to obtain a $N=2$ or $N=3$ target Laughlin state. Same loss and pump rates as in Fig.~\ref{PopsVR}. 
\label{LaughlinPop_w_at}}
\end{figure}

This physics is further illustrated in Figs.~\ref{Populations_w_at} and \ref{LaughlinPop_w_at}, where we investigate the effect on $P^{(N)}$ and $P_{\rm FQH}^{(N)}$ of shifting the emitter transition frequency $\omega_{at}$. We fix the strength of the cylindrical step potential at $V/\Delta = 50\pi$ and focus on two values of the radius $R/\ell = 3.7, 4.15$ that would yield for $\omega_{at} = \omega_0$ the highest Laughlin populations with $N = 2,3$ respectively. In the first case with $R/\ell = 3.7$ as shown in Fig.~\ref{Populations_w_at}(a), we observe that the total population $P^{(2)}$ of two particle states achieves its maximum value $~0.87$ at $\hbar(\omega_{at}-\omega_0)/\Delta \sim 0.005$ very close to the expected resonant condition, while all other $P^{(N)}$ with $N \neq 2$ remain small. In Fig.~\ref{Populations_w_at}(b) with a larger $R/\ell = 4.15$, it is seen that $P^{(2)}$ is strongly reduced to $~0.18$ on resonance, while the total population $P^{(3)}$ of three particle states is strongly enhanced and reaches a maximum value of $~0.78$. 

Similar observations follow for the populations of the Laughlin states alone as shown in Fig.~\ref{LaughlinPop_w_at}. Maximum values of the Laughlin populations are obtained in the vicinity of the resonance condition $\hbar(\omega_{at}-\omega_0)/\Delta \sim 0.001$ and in spite of some leakage to low-energy edge and quasi-hole excitations, they are still as high as $P_{\rm FQH}^{(2,3)} \approx 0.76, 0.63$ for $R/\ell = 3.7, 4.15$ respectively. 

These results in this section fully confirm our conjecture that, as long as the resonance condition $\omega_{at} \sim \omega_0$ is satisfied and the Lorentzian frequency-selectivity is small enough $\Gamma_p\ll \Delta$, one can choose to populate different $N$-particle Laughlin states with good efficiency just by changing the radius of the step potential.

\section{Spectroscopic signatures of the Laughlin state}
\label{sec:detection}
After having theoretically predicted the possibility of efficiently populating a chosen Laughlin state, we now turn to the problem of experimentally confirming the actual success of its preparation and then optically characterizing its physical properties. As a first step in this ambitious program, we propose here a relatively simple measurement of the angular-momentum-resolved emission spectrum. Our proposal is inspired by the recent work \cite{Cooper multiplicity} where RF spectroscopy was proposed as a way to extract information on a small FQH droplet of rotating bosonic atoms from the number of allowed transitions lines. In the optical case under investigation here, the role of RF spectroscopy is played by the spontaneous emission of light due to radiative photon losses through the non-perfect cavity mirrors.

\begin{table}
\begin{center}
\begin{tabular}{|c|ccccccccccc|}
 \hline 
 $m \rightarrow$  & 0 & 1 & 2 & 3 & 4 & 5 & 6 & 7 & 8 & 9 & 10 \\ 
 \hline 
 $N = 2$ & 1  & 1 & 1  & • & • & • & • & • & • & • & •  \\ 
 & (1) & (1) & (1) & • & • & • & • & • & • & • & •  \\
 \hline 
 $N = 3$ & 1 & 1  & 2  & 1  & 1 & • & • & • & • & • & •  \\ 
 & (3) & (2) & (2) & (1) & (1)& • & • & • & • & • & •  \\ 
 \hline 
 $N = 4$ & 1 & 1 & 2 & 2 & 2 & 1 & 1 & • & • & • & •  \\ 
  & (7) & (5) & (4) & (3) & (2) & (1) & (1) & • & • & • & •  \\ 
 \hline 
 $N = 5$ & 1 & 1 & 2 & 2 & 3 & 2 & 2 & 1 & 1 & • & •  \\ 
  & (15) & (11) & (9) & (6) & (5) & (3) & (2) & (1) & (1) & • & •  \\ 
 \hline 
 $N = 6$ & 1 & 1 & 2 & 2 & 3 & 3 & 3 & 2 & 2 & 1 & 1  \\
  & (30) & (23) & (18) & (13) & (10) & (7) & (5) & (3) & (2) & (1) & (1) \\
 \hline 
\end{tabular} 
\caption{Multiplicities $\bar{\mathcal{N}}_f$ of allowed transitions with angular momentum $m\hbar$ from the $N$-particle Laughlin state. Below each multiplicity, the total number $\mathcal{N}_f$ of possible final states within the $(N-1)$ Laughlin manifold is written in parantheses. \label{multiplicities}} 
\end{center}
\end{table}

Supposing that the initial state of our fluid is a pure $N$-particle Laughlin state, when a cavity photon of angular momentum $m\hbar$ is lost, the resulting many-particle state belongs to the $(N-1)$-particle Laughlin manifold and can involve edge or quasi-hole excitations of the $(N-1)$-particle Laughlin state (see Appendix C). The wave function of this state can be written as a product of the $(N-1)$-particle Laughlin wave function and a symmetric polynomial in coordinates $\{z_1,\ldots,z_{N-1}\}$ with total power $2(N-1)-m$. Therefore, the number of such final states $\mathcal{N}_f$ in the $(N-1)$-particle Laughlin manifold for a given $m$ can simply be determined by the number $\mathcal{N}_f$ of ways to distribute $[2(N-1)-m]\hbar$ angular momenta among $N-1$ indistinguishable particles. In formal terms, we are looking for a (integer) partition $\lambda$ of the number $2(N-1)-m$ of maximum length $N-1$.

However, as it was pointed out in~\cite{Cooper multiplicity} not all these transitions have a significant matrix element. Losing a photon from a Laughlin state can in fact be seen as resulting in the creation of two quasi-holes, which imposes the constraint that the largest single-particle angular momentum in the distribution of the extra angular momentum be $2\hbar$. The same condition can be obtained noting that the largest single-particle angular momentum in the final $(N-1)$-particle state cannot exceed that of the initial Laughlin state with $N$ particles. As discussed in detail in~\cite{Cooper multiplicity}, this many-body selection rule is not exact; still the matrix elements for non-allowed transitions are orders of magnitude smaller and practically negligible.

The selection rule can be put in formal terms by requiring that the largest entry $\lambda_1$ of the partition $\lambda$ be at most $2$. This significantly reduces the number of allowed transitions to the number $\bar{N}_f$ of (integer) partition $\lambda$ of the number $2(N-1)-m$ of maximum length $N-1$ and of maximal entry $2$.

As a very important side remark, it is interesting to note the transition to the $(N-1)$-particle Laughlin state is possible according to this criterion and therefore has a sizable matrix element: as the matrix element is the same also for decay and for pumping, this guarantees the efficiency of pumping upwards on this transition.

In Table \ref{multiplicities}, we show both $\mathcal{N}_f$ and $\bar{\mathcal{N}}_f$ for transitions starting from the $N$-particle Laughlin state with $N = 2,\ldots,6$ for a given $m$: remarkably, $\mathcal{N}_f$ greatly exceeds $\bar{\mathcal{N}}_f$ especially for small $m$ and shows a highly symmetric and unique pattern as a function of $m$. As a result, an experimental counting of the number of allowed transitions should be a compelling evidence for the Laughlin nature of the generated state. 

In order to determine $\bar{\mathcal{N}}_f$ experimentally, one first needs to lift the degeneracies and this is naturally done in our setup in the presence of the external potential. Multiplicities of the allowed transitions can then be inferred by measuring the angular-momentum-resolved emission spectrum. As usual in quantum optics, the emission spectrum can be extracted either directly via a suitably selective spectrometer or from the Fourier transform of the first-order correlation function $g^{(1)}_m(\tau)\equiv {\rm Tr}[a^{\dagger}_m(\tau)a_m\rho_{ss}]$, where $a_m$ annihilates a photon with $m$ units of angular momentum and $a^{\dagger}_m(\tau)$ is the time-evolved operator after a time $\tau$~\cite{Zoller,PRA_back}. Light filtering according to the angular momentum can be performed using, e.g., holograms \cite{angular momentum}. 

Under the assumption that $\Gamma_e,\Gamma_l$ are smaller than the typical energy scale of the Hamiltonian evolution, we can take for simplicity $a^{\dagger}_m(\tau) \simeq e^{i\mathcal{H}\tau/\hbar}a^{\dagger}_me^{-i\mathcal{H}\tau/\hbar}$. Expressing $g^{(1)}_m(\tau)$ in the eigenstate basis of $\mathcal{H}$ as $g^{(1)}_m(\tau) = \sum_{i,f} e^{i(\omega_i-\omega_f)\tau}\langle i|a^{\dagger}_m|f\rangle\langle f|a_m\rho_{ss}|i\rangle $, where $|i\rangle$ ($|f\rangle$) is the initial (final) state, we define the Fourier amplitude $\mathcal{F}_{i\rightarrow f}^{(m)}\equiv \langle i|a^{\dagger}_m|f\rangle\langle f|a_m\rho_{ss}|i\rangle$ corresponding to the transition frequency $\omega_{if} \equiv \omega_i-\omega_f$, which yields $g^{(1)}_m(\tau)= \sum_{i,f} e^{i \omega_{if}\tau} \mathcal{F}_{i\rightarrow f}^{(m)}$ and a series of corresponding peaks of frequencies $\omega_{if}$ and strength $\mathcal{F}_{i\rightarrow f}^{(m)}$ in the emission spectrum. The multiplicity of allowed transitions corresponding to each $m$ can then be found by counting the dominant peaks in the emission spectrum.

In Figs.~\ref{FourierN2to1}(d) and \ref{FourierN3to2}(d) we display sketches of the emission spectra as a function of $\omega_{if}-\omega_0$ with peak intensities $\mathcal{F}_{i\rightarrow f}^{(m)}$. The two figures correspond to configurations where the target states to be populated are the $N = 2$ and $N = 3$ Laughlin states, respectively. In the upper (a,b,c) panels of these figures, we also show the energy eigenvalues $E^{\prime}_N = E_N-N\hbar\omega_0$ of the isolated system scaled by the Laughlin gap $\Delta$ as a function of total angular momentum $L_z$, focusing on the low-energy sectors of each relevant $N$-particle manifold. 

Using $R/\ell = 3.7$ to get a high $N = 2$ Laughlin population, we expect three dominant peaks around $\omega_{if} \sim \omega_0$ one for each $m = 0,1,2$ corresponding to the transition from the $N = 2$ Laughlin state with total angular momentum $L_z = 2\hbar$ to the three states in the $N = 1$ manifold with angular momentum $L_z\le2\hbar$ (see Table \ref{multiplicities}). This is what we see in Fig.~\ref{FourierN2to1}(d). Besides the dominant peaks appearing at the expected values of the transition frequency as indicated by vertical dashed lines, there are smaller peaks which correspond to other transitions as the initial state is not a pure Laughlin state ($P^{(2)}_{\rm FQH} \approx 0.76$). 

When the target is the $N = 3$ Laughlin state with $R/\ell = 4.15$, we expect six dominant peaks (see Table \ref{multiplicities}), one for $m = 0,1,3,4$ and two for $m = 2$. Fig.~\ref{FourierN3to2}(d) confirms this conjecture as the peaks corresponding to these transitions are still well stronger than other spurious peaks due to the siginificantly non-perfect preparation of the $N = 3$ Laughlin state (of population $P^{(3)}_{\rm FQH} \approx 0.63$) and to the non-allowed transitions.

As a final remark, it is interesting to note that an efficient preparation of the $N$-particle Laughlin state requires that the energy shift of the $(N+1)$-particle Laughlin state and the separation from quasi-hole and edge excitations of the $N$-particle Laughlin manifold be larger than the fast pumping rate $\Gamma_p$. On the other hand, spectral isolation of the allowed emission lines from the target Laughlin state to the $(N-1)$-particle Laughlin manifold only requires that the spectral separation be larger than the linewidth, which is on the order of $\Gamma_{e,l}$. In our frequency-dependent incoherent pumping scheme, these latter are assumed from the outset to be smaller than $\Gamma_p$, which somehow compensates for the fact that the spectral splittings due to confinement are typically smaller in the $(N-1)$-particle manifold.

\section{CONCLUSIONS}
\label{sec:conclu}
In this work we have studied the possibility of using a frequency-selective incoherent pumping mechanism to create and detect a strongly-correlated fractional quantum Hall droplet of effectively interacting photons with a well-defined particle number. Even though our theory is focused on the non-planar ring cavity configuration, where Landau levels for photons and Rydberg-EIT photon blockade have been recently observed~\cite{synthetic Landau levels,blockade}, the ideas presented here are fully general and may be extended to other optical configurations combining a synthetic magnetic field for photons and strong photon-photon interactions.

Exploiting the equal separation between the energies of any two Laughlin states with consecutive number of particles and using a suitably designed external potential to set a cutoff for the particle number, we have shown that a particular $N$-particle Laughlin state can be selectively populated. The incompressibility of Laughlin states provides a sizable blue-shift of all edge and quasi-hole excitations of the $N$-particle Laughlin state and of the whole $(N+1)$-particle manifold, which effectively prevents the actual population of all these states by the frequency-dependent incoherent pump. 

Based on the microscopic structure of Laughlin states and their excitations, we have also proposed a method to experimentally get unambiguous signatures of Laughlin physics in our setup. The proposed method employs a standard spectral measurement of the emitted light along with an angular momentum selection process. We showed that if the population of a certain Laughlin state dominates over all other populations, the spectral distribution of the emission yields clear peaks, whose number and position is determined by the allowed transitions to states in the Laughlin manifold and carries unambiguous information on the Haldane fractional exclusion statistics of quantum Hall states.

While this paper has reported the general idea of the pumping scheme and has verified its efficiency for quite idealized systems parameters, future work will investigate more sophisticated potentials showing larger confinement-induced gaps, so as to relax the constraint on the atomic linewidth parameters. At the same time, we are also exploring the possibility of tayloring the atomic spatial distribution to further improve the fidelity of the Laughlin state preparation.

\section*{ACKNOWLEDGMENTS}

We are grateful to Elia Macaluso for continuous fruitful collaboration on Laughlin physics and, in particular, for bringing Ref. \cite{Cooper multiplicity} to our attention. Insightful exchanges with Mohammad Hafezi and Jonathan Simon are also warmly acknowledged. This work was supported by the EU-FET Proactive grant AQuS, Project No. 640800, and by the Autonomous Province of Trento, partially through the project ``On silicon chip quantum optics for quantum computing and secure communications" (``SiQuro").  IC acknowledges the Kavli Institute for Theoretical Physics, University of California, Santa Barbara (USA) for the hospitality and support during the early stage of this work.

\section*{APPENDIX A: LAUGHLIN STATE POPULATION AS A FUNCTION OF PUMP PARAMETERS}
\label{app:pump parameters}

\begin{figure}[htbp]
\includegraphics[scale=0.08]{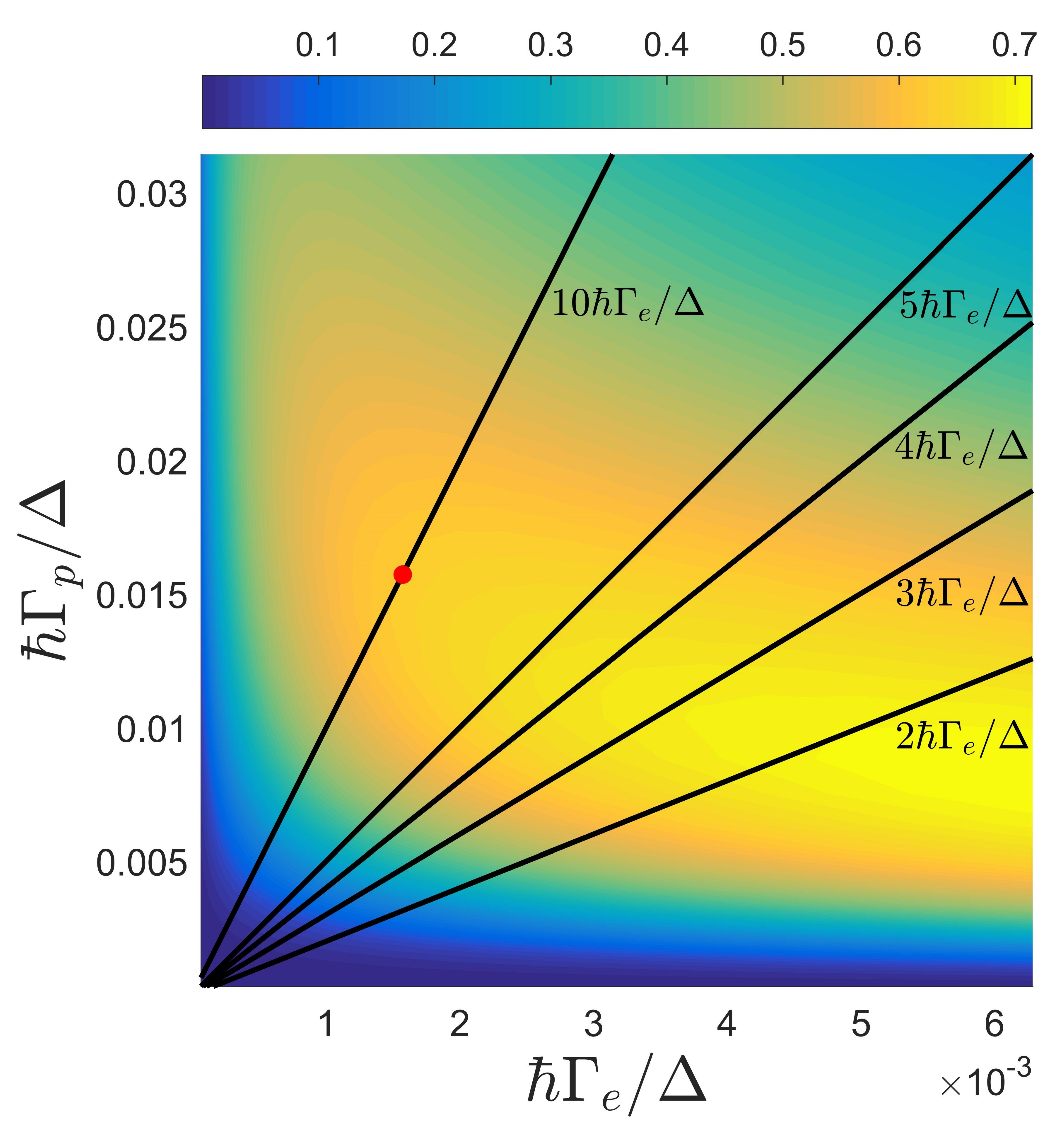}
\caption{Population $P^{(3)}_{\rm FQH}$ of the $N = 3$ Laughlin state for $V/\Delta = 50\pi$, $R/\ell = 4.15$, and $\omega_{at}=\omega_0$ as a function of pump $\Gamma_p$ and emission $\Gamma_e$ rates for a fixed loss rate $\hbar\Gamma_l/\Delta = \pi\times10^{-5}$. Black lines correspond to several fixed $\Gamma_p/\Gamma_e = 2,3,4,5,10$ ratios and the red dot locates the values $\hbar\Gamma_e/\Delta = 5\pi\times 10^{-4}$ and $\hbar\Gamma_p/\Delta = 5\pi\times 10^{-3}$ used in the main text.
\label{LaughlinN3Pop_parameters}}
\end{figure}

Fig.~\ref{LaughlinN3Pop_parameters} shows the $N = 3$-particle Laughlin state population $P^{(3)}_{\rm FQH}$ as a function of $\Gamma_p$ and $\Gamma_e$ for fixed $\hbar\Gamma_l/\Delta = \pi\times10^{-5}$ and $\omega_{at} = \omega_0$ in the presence of a step potential with radius $R/\ell = 4.15$. Although the maximum value of $P^{(3)}_{\rm FQH}$ in the parameter range we explored is $\approx 0.71$, somewhat different parameters $\hbar\Gamma_e/\Delta = 5\pi\times10^{-4}$ and $\Gamma_p = 10\Gamma_e$ were chosen for the calculations presented in the main text, yielding $P^{(3)}_{\rm FQH} \approx 0.63$.

These values were chosen to have $\Gamma_e$ significantly smaller than $\Gamma_p$, so as to be consistent with the approximations underlying the derivation of the generalized master equation discussed in Section \ref{subsec:Losses and Incoherent Pumping}.

\section*{APPENDIX B: DIFFERENT EXTERNAL POTENTIAL PROFILES FOR LARGER POPULATIONS}
\label{app:potential}

\begin{figure}[htbp]
\includegraphics[scale=0.065]{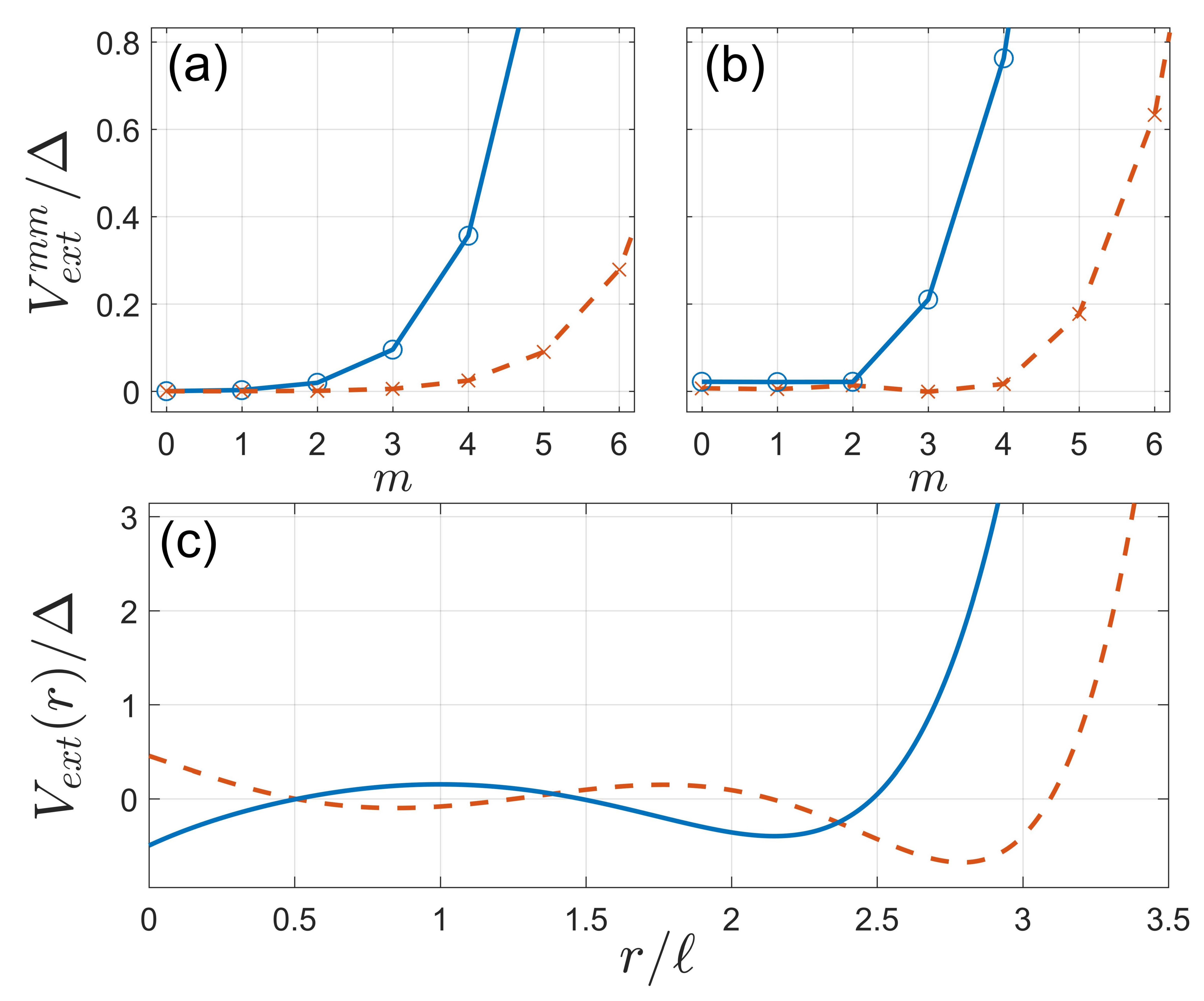}
\caption{(a) Matrix elements $V_{ext}^{mm}$ of the cylindrically-symmetric step potential with $V/\Delta = 50\pi$ in the angular momentum basis, for $R/\ell = 3.7$ (solid blue line) and $R/\ell = 4.15$ (red dashed line). (b) Matrix elements of two external potential profiles given in (c), one yielding $P^{(2)}_{\rm FQH} \approx 0.91$ (solid blue line) and the other giving $P^{(3)}_{\rm FQH} \approx 0.76$ (red dashed line) for the same loss and pump parameters as in Fig.~\ref{PopsVR}. Note that both potentials keep monotonically increasing for large $r$.
\label{potential_matrix_elements}}
\end{figure}

\begin{figure}[htbp]
\includegraphics[scale=0.46]{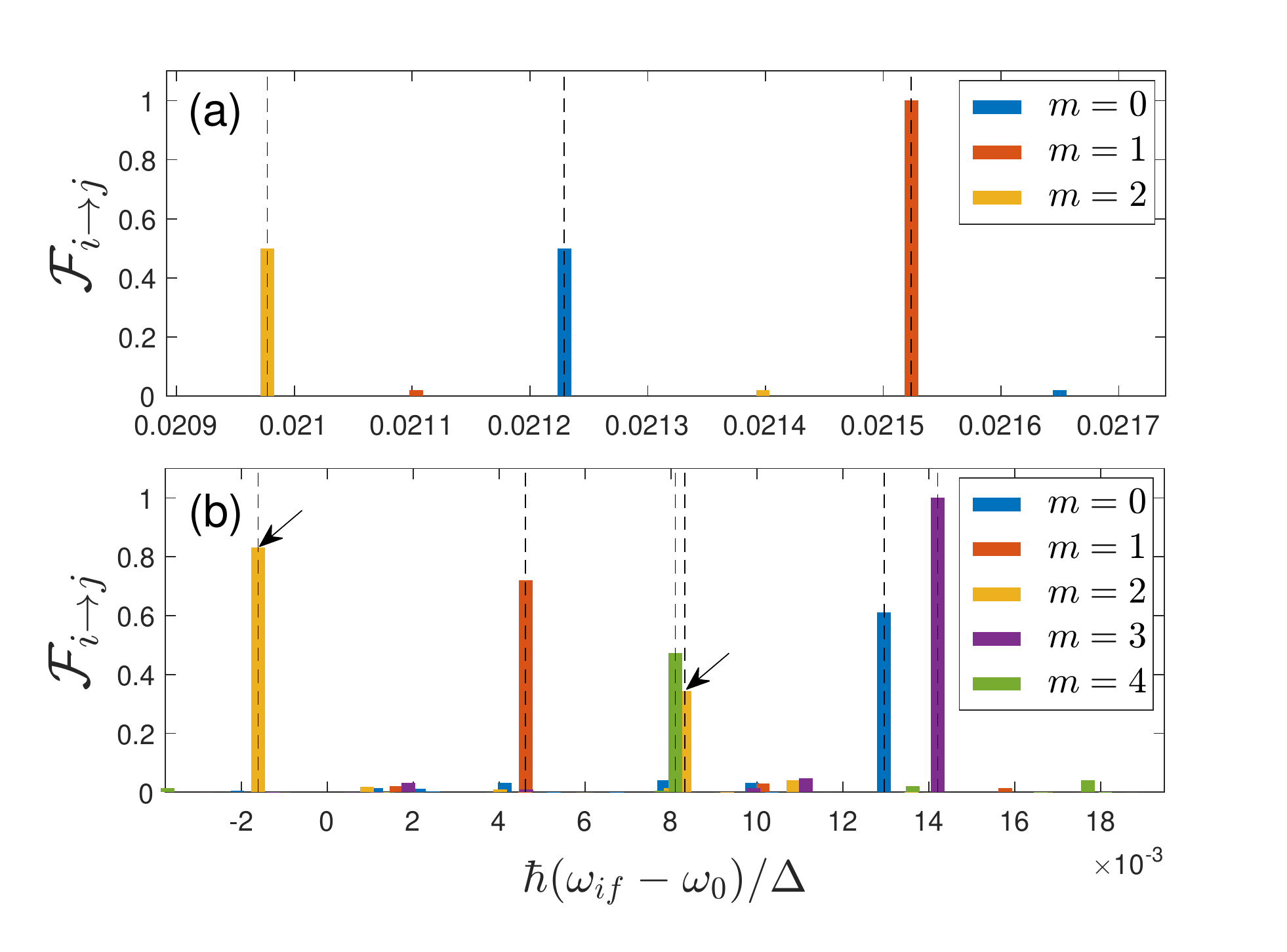}
\caption{Scaled Fourier amplitudes $\mathcal{F}_{i\rightarrow f}^{(m)}$ of transitions as a function of the difference between the lost-photon frequency $\omega_{if} = \omega_i-\omega_f$ and the reference frequency $\omega_0$ for the potential profiles of panel (c) in Fig.~\ref{potential_matrix_elements}, (a) shown by a solid blue line and (b) by a red dashed line. Different colors designate the angular momentum $m\hbar$ of the photon lost from the cavity. Vertical dashed lines show the position of allowed transitions. Transitions to two states involving the loss of $2\hbar$ angular momenta are indicated by arrows in (b). Same loss and pump parameters as in Fig.~\ref{PopsVR}.
\label{FourierN2to1and3to2}}
\end{figure}

In this Appendix, we show that a potential profile different from the step one might yield a slightly larger population of a target Laughlin state. Among the many potential profiles we explored, we present two of them which maintain cylindrical symmetry but have the property that the matrix elements of the potential in the angular momentum basis $V_{ext}^{mm}$ more rapidly increase for $m>2(N-1)$ [Fig.~\ref{potential_matrix_elements}(b)] as compared to the matrix elements of the step potential [Fig.~\ref{potential_matrix_elements}(a)]. These more sophisticated potentials are shown in Fig.~\ref{potential_matrix_elements}(c) and are optimized for respectively $N=2$ (solid line) and $N=3)$ (dashed line). In the language of \cite{elia}, these potentials go in the direction of the steep hard wall limit.

Interestingly, the potential profile shown as a solid line in Fig.~\ref{potential_matrix_elements}(c) gives $P^{(2)}_{\rm FQH} \approx 0.91$ and the one shown as the dashed line yields $P^{(3)}_{\rm FQH} \approx 0.76$: using the more sophisticated potentials, one obtains values that are of course better than the simple step potential, but the improvement does not appear to be dramatic. In Fig.~\ref{FourierN2to1and3to2}(a) and (b) we display the intensity of the emission lines for the more sophisticated potentials: compared to the step potential case illustrated in Figs.~\ref{FourierN2to1} and \ref{FourierN3to2}, the larger population of the Laughlin state also reflects in that the emission peaks corresponding to the expected transitions dominate in a clearer way the ones due to other transitions.

\section*{APPENDIX C: LOSS OF A PARTICLE FROM THE LAUGHLIN STATE}
\label{app:loss from Laughlin}

In this last Appendix we reproduce for completeness the results from Appendix B of our previous work \cite{coherent Laughlin 3}.

When a particle with angular momentum $m \hbar$ is annihilated from a generic bosonic $N$-particle state $|\Phi_N\rangle$ lying in the LLL, the state of the system can be found by applying $a_m = \int dz^{\prime}dz^{\prime\ast} \varphi^{\ast}_m(z^{\prime})\Psi(z^{\prime})$ to $|\Phi_N\rangle = \int dz_1dz_1^{\ast}\ldots dz_Ndz_N^{\ast} \Phi_N(z_1,\ldots,z_N)\Psi^{\dagger}(z_1)\ldots \Psi^{\dagger}(z_N)|{\rm vac.}\rangle$, where $|{\rm vac.}\rangle$ is the vacuum state. The resultant state is

\begin{multline}
|\Phi_{N-1}^{\prime}\rangle \equiv a_m|\Phi_N\rangle = N \int dz_1\,dz_1^{\ast}\ldots dz_{N-1}\,dz_{N-1}^{\ast} \\ 
\times \left[ \int \Phi_N(z_1,\ldots,z_{N-1},z)\varphi^{\ast}_m(z)dz\, dz^{\ast}\right] \\
\times \Psi^{\dagger}(z_1)\ldots \Psi^{\dagger}(z_{N-1})|{\rm vac.}\rangle.
\end{multline}   
The wave function corresponding to this state with $N-1$ particles is identified as $\Phi_{N-1}^{\prime}(z_1,\ldots,z_{N-1}) = \int \Phi_N(z_1,\ldots,z_{N-1},z)\varphi^{\ast}_m(z)\,dz\, dz^{\ast}$ up to a normalization constant. Choosing $z = z_N$ and noting that $\varphi^{\ast}_m(z_N) \propto z_N^{\ast m}$, $\Phi_{N-1}^{\prime}$ is found to be proportional to the multinomial term multiplying $z_N^{m}$ in $\Phi_N$ as it is the only surviving term in the integral expression for $\Phi_{N-1}^{\prime}$. 

For the Laughlin wave function (\ref{WF_FQH}) it can be seen that $\Phi_{N-1}^{\prime}$ is the $(N-1)$-particle Laughlin wave function times a symmetric polynomial in coordinates $\{z_1,\ldots,z_{N-1}\}$ with total power $2(N-1)-m$.


\begin{thebibliography}{99}
\bibitem{superoperator}
We define a superoperator $\mathcal{S}$ in terms of which the master equation can
be written as $\partial\tilde{\rho}/\partial t = \mathcal{S}\tilde{\rho}$, where $\tilde{\rho}$ is a vector constructed
from the elements of the matrix representation of $\rho$. Finding the steady-state density matrix $\rho_{ss}$ then reduces to finding the eigenvector of $\mathcal{S}$ with zero eigenvalue.
\end{thebibliography}
\end{document}